# Learning Universal Adversarial Perturbations with Generative Models

Jamie Hayes and George Danezis
University College London
j.hayes@cs.ucl.ac.uk
g.danezis@ucl.ac.uk

***Abstract*—Neural networks are known to be vulnerable to *adversarial examples*, inputs that have been intentionally perturbed to remain visually similar to the source input, but cause a misclassification. It was recently shown that given a dataset and classifier, there exists so called *universal adversarial perturbations*, a single perturbation that causes a misclassification when applied to any input. In this work, we introduce *universal adversarial networks*, a generative network that is capable of fooling a target classifier when it's generated output is added to a clean sample from a dataset. We show that this technique improves on known universal adversarial attacks.**

## I. Introduction

Machine Learning models are increasingly relied upon for safety and business critical tasks such as in medicine [22], [30], [41], robotics and automotive [28], [32], [40], security [2], [17], [38] and financial [13], [18], [36] applications. Recent research shows that machine learning models trained on entirely uncorrupted data, are still vulnerable to *adversarial examples* [7], [12], [23], [24], [35], [37]: samples that have been maliciously altered so as to be misclassified by a *target* model while appearing unaltered to the human eye.

Most work has focused on generating perturbations that cause a *specific* input to be misclassified, however, it has been shown that adversarial perturbations generalize across many inputs [35]. Moosavi-Dezfooli *et al.* [19] showed, in the most extreme case, that given a target model and a dataset, it is possible to construct a single perturbation that when applied to *any* input, will cause a misclassification with high likelihood. These are referred to as *universal adversarial perturbations* (UAPs).

In this work, we study the capacity for generative models to learn to craft UAPs on image datasets, we refer to these networks as *universal adversarial networks* (UANs). We show that a UAN is able to sample from noise and generate a perturbation such that when applied to *any* input from the dataset, it will result in a misclassification in the target model. Furthermore, we show perturbations produced by UANs: improve on state-of-the-art methods for crafting UAPs (Section IV-A), have robust transferable properties (Section IV-D), and reduce the success of recently proposed defenses [1] (Section V).

## II. Background

We define adversarial examples and UAPs along with some terminology and notation. We then introduce the threat model considered, and the datasets we use to evaluate the attack.

### A. Adversarial Examples

Szegedy *et al.* [35] casts the construction of adversarial examples as an optimization problem. Given a *target model*, $f$, and a *source* input $x$, which is classified correctly by $f$ as $c$, the attacker aims to find a perturbation, $\delta$, such that $x + \delta$ is perceptually identical to $x$ but $f(x + \delta) \neq c$. The attacker tries to minimize the distance between the source image and adversarial image under an appropriate measure. The problem space can be framed to find a specific misclassification in a *targeted* attack, or *any* misclassification, referred to as a *non-targeted* attack.

In the absence of a distance measure that accurately captures the perceptual differences between a source and adversarial image, the $\ell_p$ metric is usually minimized [35]. Related work commonly uses the $\ell_2$ and $\ell_\infty$ metrics [3], [4], [6], [10], [14], [16], [19], [20], [42]. The $\ell_2$ metric measures the Euclidean distance between two images, while the $\ell_\infty$ metric measures the largest pixel-wise difference between two images (Chebyshev distance). We follow this practice here and construct attacks optimizing under both metrics.

A UAP is an adversarial perturbation that is independent of the source image. Given a *target model*, $f$, and a dataset, $X$, a UAP is a perturbation, $\delta$, such that $\forall x \in X$, $x + \delta$ is a valid input and $\Pr(f(x + \delta) \neq f(x)) = 1 - \tau$, where $0 < \tau << 1$.

### B. Threat Model

We consider an attacker whose goal is to craft UAPs against a target model, $f$. The adversarial image constructed by the attacker should be visually indistinguishable to a source image, evaluated through either the $\ell_2$ or $\ell_\infty$ metric.

Our attacks assume white-box access to $f$, as we backpropagate the error of the target model back to the UAN. In line with related work on UAPs [19], we consider a *worst-case* scenario with respect to data access, assuming that the attacker has knowledge of, and shares access to, any training data samples. We will not discuss the real-world limitations of that assumption here, but will follow that practice.

### C. Datasets

We evaluate attacks using two popular datasets in adversarial examples research, CIFAR-10 [15] and ImageNet [29].

The CIFAR-10 dataset consists of 60,000, 32×32 RGB images of different objects in ten classes: airplane, automobile, bird, cat, deer, dog, frog, horse, ship, truck. This is split into 50,000 training images and 10,000 validation images.

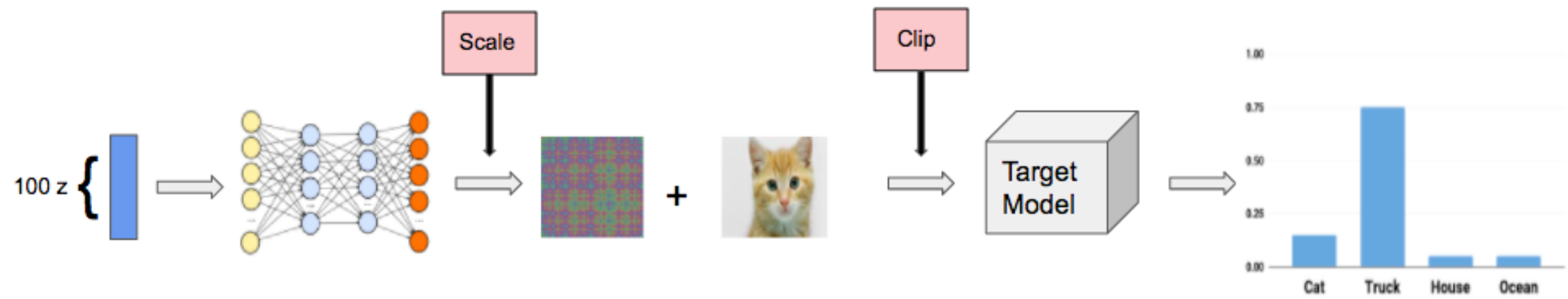

Fig. 1: Overview of the attack. A random sample from a normal distribution is fed into a UAN. This outputs a perturbation, which is then scaled and added to an image. The new image is then clipped and fed into the target model.

Our pre-trained models: VGG-19 [31], ResNet-101 [9], and DenseNet [11], used as the target models, score 91.19%, 93.75%, and 95.00% test accuracy, respectively. State-of-the-art models on CIFAR-10 are approximately 95% accurate.

We use the validation dataset of ImageNet, which consists of 50,000 RGB images, scaled to 224×224. The images contain 1,000 classes. The 50,000 images are split into 40,000 training set images and 10,000 validation set images. We ensure classes are balanced, such that any class contains 40 images in the training set and 10 images in the validation set. Our pre-trained models: VGG-19 [31], ResNet-152 [9], and Inception-V3 [34], used as the target models, score 71.03%, 78.40%, and 77.22% top-1 test accuracy, respectively.

## III. Universal Adversarial Networks

### A. Attack Description

An overview of the attack is given in Figure 1. Let a UAN model be denoted by $\mathcal{U}$, and a target model by $f$. $\mathcal{U}$ takes as input a vector, $z$, sampled from a normal distribution $\mathcal{N}(0,1)^{100}$, and outputs a perturbation, $\delta$. This is then scaled by a factor $\omega \in (0, \frac{\epsilon}{\|\delta\|_p}]$, where $\epsilon$ is the maximum permitted perturbation and $p = 2$ or $\infty$. In practice, we start with a small $\omega$ (e.g. $\omega = \frac{\epsilon}{10\cdot\|\delta\|_p}$) and increment this value whenever the training loss plateaus. The scaled perturbation $\delta' = \omega \cdot \delta$, is added to an image $x$ from a dataset $X$, to produce an adversarial image. This is then clipped into the target model's input range before being fed into the target model, $f$, which outputs a probability vector, $\rho$ [1]. If $\arg\max_i f(x) \neq \arg\max_i f(\delta' + x)$, a successful adversarial example has been found. Since $\mathcal{U}(z)$ is not conditioned on any image in the dataset, $\mathcal{U}$ learns how to construct image independent adversarial perturbations, namely universal adversarial perturbations.

Given an input $x \in X$, let the class label predicted by $f$ be $c_0$. For non-targeted attacks, *any* misclassification in the target model suffices, thus, the non-targeted attack aims to maximize the most probable predicted class other than $c_0$. Our non-targeted loss function is adapted from works by Carlini and Wagner [4] and Chen *et al.* [5], and is given by:

$$L_{nt} = \underbrace{\log[f(\delta'+x)]_{c_0} - \max_{i \neq c_0} \log[f(\delta'+x)]_i}_{L_{fs}} + \underbrace{\alpha \cdot \left\|\delta'\right\|_p}_{L_{dist}} \quad (1)$$

[1]If $f$ outputs logits instead of a probability vector, we take the softmax of the logits.

The first term in (1), $L_{fs}$, is minimized when the adversarial predicted class is not $c_0$. This is adapted from the Carlini and Wagner loss function [4] that introduces a confidence threshold, $\kappa$. If we want universal adversarial perturbations that cause misclassifications with high confidence, we stop minimizing only when:

$$\kappa > \max_{i \neq c_0} \log[f(\delta'+x)]_i - \log[f(\delta'+x)]_{c_0}$$

In specifying a confidence threshold for adversarial examples, (1) becomes:

$$L_{nt} = \max\{\log[f(\delta'+x)]_{c_0} - \max_{i \neq c_0} \log[f(\delta'+x)]_i, -\kappa\} + \alpha \cdot \left\|\delta'\right\|_p \quad (2)$$

In all experiments we set $\kappa = 0$, and so stop optimizing once an adversarial example is found. To minimize the perturbation applied to an image, $L_{fs}$ is summed with a distance loss, $L_{dist} = \alpha \cdot \|\delta'\|_p$, where $\alpha \in \mathbb{R}^+$; this minimizes the norm of the universal adversarial perturbation. The logarithmic term in $L_{fs}$ is necessary since most target models have a skewed probability distribution, with one class prediction dominating all others, thus the logarithmic term reduces the effect of this dominance.

For a targeted attack, we compute a universal adversarial perturbation that transforms *any* image to a chosen class, $c$. Under this setting, we optimize using the follow loss function:

$$L_t = \max\{\max_{i \neq c} \log[f(\delta'+x)]_i - \log[f(\delta'+x)]_c, -\kappa\} + \alpha \cdot \left\|\delta'\right\|_p, \quad (3)$$

The full description of the UAN model is given in Table I and hyperparameters used in experiments are given in Table II. We define the relative perturbation, $\zeta_p = \frac{\|\delta'\|_p}{\|x\|_p}$; the value of the norm of $\delta'$ over the norm of the original image, $x$. We set $\zeta_p = 0.04$ in all experiments [2] [3]. For all experiments in Section IV, we report the *error rate* of the target model on adversarial images; a perfect attack would achieve an error rate of 1.00, while a perfect classifier achieves an error rate of 0.00.

[2]Code available at https://github.com/jhayes14/UAN

[3]Note, this is equivalent to the experimental settings in Moosavi-Dezfooli *et al.* [19] of $\epsilon = 10$ for $p = \infty$, and $\epsilon = 2000$ for $p = 2$.

TABLE I: UAN model architecture. $IS$ refers to the image size: 32 for CIFAR-10 experiments and 224 for ImageNet experiments.

| Layer | Shape |
|---|---|
| Input | 100 |
| Deconv + Batch Norm + ReLU | $256 \times 3 \times 3$ |
| Deconv + Batch Norm + ReLU | $128 \times 5 \times 5$ |
| Deconv + Batch Norm + ReLU | $64 \times 9 \times 9$ |
| Deconv + Batch Norm + ReLU | $32 \times 17 \times 17$ |
| Deconv + Batch Norm + ReLU | $3 \times 33 \times 33$ |
| FC + Batch Norm + ReLU | 512 |
| FC + Batch Norm + ReLU | 1024 |
| FC | $3 \times IS \times IS$ |

TABLE II: UAN hyperparameters.

| Parameter | Dataset | |
|---|---|---|
| | CIFAR-10 | ImageNet |
| Learning Rate | $2 \cdot 10^{-4}$ | $2 \cdot 10^{-4}$ |
| Beta 1 | 0.5 | 0.5 |
| Beta 2 | 0.999 | 0.999 |
| Batch Size | 128 | 64 |
| Epochs | 500 | 150 |
| $\ell_p$ loss weight ($\alpha$) | 4.0 | 4.0 |

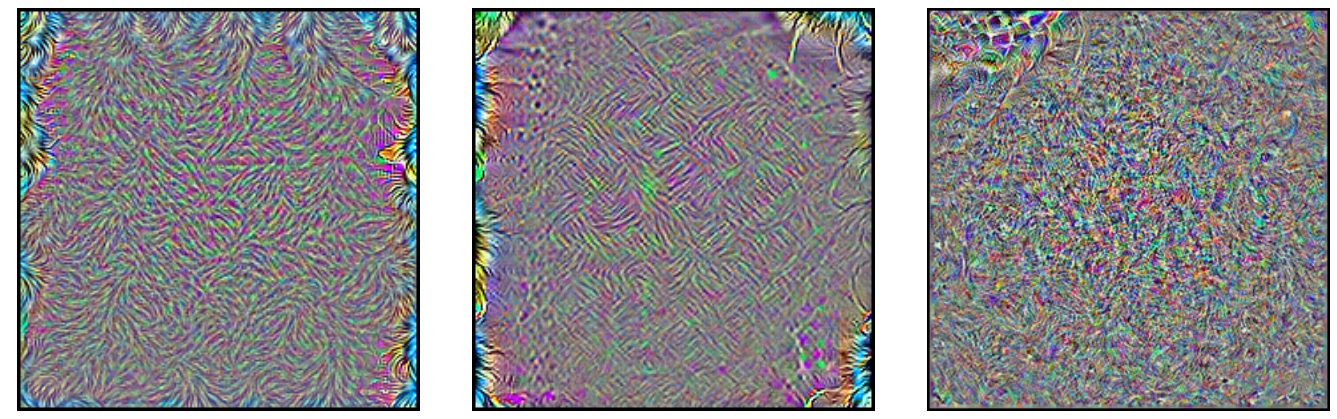

(a) VGG-19 (b) ResNet-152 (c) Inception-V3

Fig. 2: UAPs generated by a UAN for ImageNet.

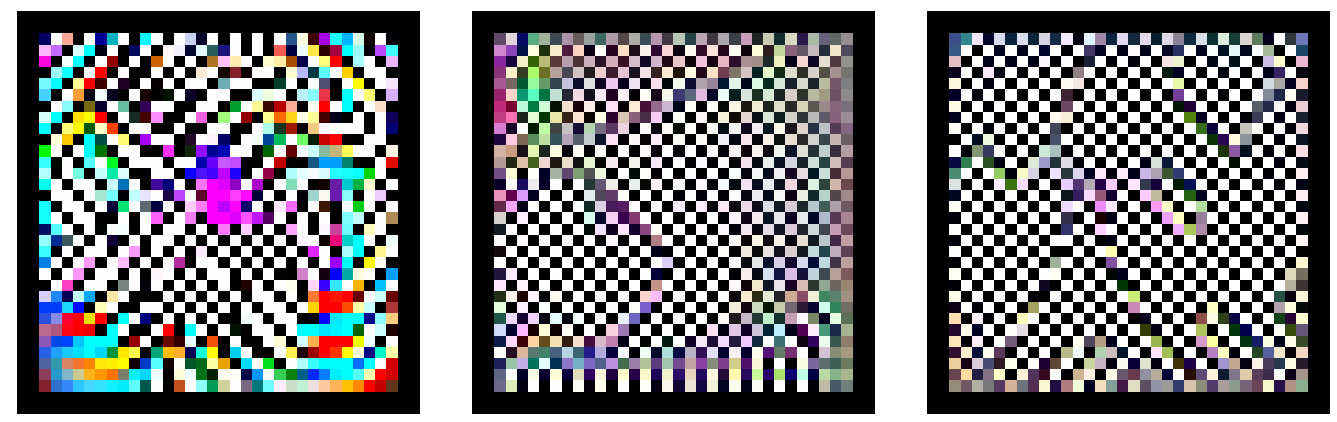

(a) VGG-19 (b) ResNet-101 (c) DenseNet

Fig. 3: UAPs generated by a UAN for CIFAR-10.

# IV. Evaluation

## A. Comparison with previous work

We now compare our method for crafting UAPs with two state-of-the-art methods:

- Moosavi-Dezfooli *et al.* [19] constructs a UAP iteratively; at each step an input is combined with the current constructed UAP, if the combination does not fool the target model, a new perturbation with minimal norm is found that does fool the target model. The attack terminates when a threshold error rate is met.
- Mopuri *et al.* [21] develop a method for finding a UAP for a target model that is independent of the dataset. They construct a UAP by first starting with random noise and iteratively update it to over-saturate features learned at successive layers in the target model, causing neurons at each layer to output useless information to cause the desired misclassification. They optimize the UAP by adjusting it with respect to the loss term:

$$L = -\log(\prod_{i=1}^{K} \bar{l}_i(\delta)), \text{ such that } ||\delta||_\infty < \gamma,$$

where, $\bar{l}_i(\delta)$ is the average of the output at layer $i$ for perturbation $\delta$, and $\gamma$ is the maximum permitted perturbation.

Table III compares our UAN method of generating UAPs against the two attacks described above for both CIFAR-10 and ImageNet, in a non-targeted attack setting. We consistently outperform both attack methods. UAPs for the ImageNet and CIFAR-10 datasets are given in Figure 2 and Figure 3, respectively. A selection of adversarial images for the ImageNet dataset is given in Figure 9.

## B. Transferability

An adversarial image is *transferable* if it successfully fools a model that was not its original target. Transferability is a yardstick for the robustness of adversarial examples, and is the main property used by Papernot *et al.* [23], [24] to construct black-box adversarial examples. They construct a white-box attack on a local target model that has been trained to replicate the intended target models decision boundaries, and show that the adversarial examples can successfully transfer to fool the black-box target model.

To measure the transferability properties of perturbations crafted by a UAN, we create 10,000 adversarial images (constructed via the $\ell_\infty$ metric) - one for each image in the CIFAR-10 validation set - and apply them to a target model that was not used to train the UAN. Table IV presents results for transferability of a non-targeted attack on three target models - VGG-19, ResNet-101, and DenseNet. We find that UAPs crafted using a UAN do transfer to other models. For example, a UAN trained on VGG-19, and evaluated on ResNet-101, the error rate is 61.2%, a drop of just 5.4% from evaluating on the original target model (VGG-19).

We also measure the capacity for a UAN to learn to fool an ensemble of target models. We trained a UAN against VGG-19, ResNet-101, and DenseNet, simultaneously, on CIFAR-10, where the UAN loss function is a linear combination of the losses of each target model. From Table IV, we see that a UAN trained against an ensemble of target models is able to fool at comparable rates to single target models.

## C. Generalizability

Moosavi-Dezfooli *et al.* [19] have shown that UAPs are not unique; there exists many candidates that perform equally well against a target model. If a UAN is truly modeling the distribution of UAPs the output should not be unique. In Figure 5, we measure the MSE (mean square error) and SSIM (structural similarity index) [39] of $\mathcal{U}(z_1)$,$\mathcal{U}(z_2)$ for $z_1, z_2 \leftarrow \mathcal{N}(0,1)^{100}$, $z_1 \neq z_2$, at successive training steps, for the ImageNet dataset. Since we expect a high degree of structure in a UAP, SSIM is measured in addition to MSE, as it has been argued that MSE does not map well to a human's perception of image structure [25], [39].

TABLE III: Comparison of error rates for UAN against Moosavi-Dezfooli *et al.* [19] and Mopuri *et al.* [21]. Note that the Mopuri *et al.* [21] method for crafting UAPs is only optimized under the $\ell_\infty$ metric. We set $\zeta_p = 0.04$, this is equivalent to $\epsilon = 2000$ for an $\ell_2$ attack and $\epsilon = 10$ for an $\ell_\infty$ attack.

| **Metric** | **Attack** | | **CIFAR-10** | | | **ImageNet** | | |
|---|---|---|---|---|---|---|---|---|
| | | | VGG-19 | ResNet-101 | DenseNet | VGG-19 | ResNet-152 | Inception-V3 |
| $\ell_2$ | UAN | *Train* | **0.689** | **0.861** | 0.753 | 0.889 | **0.918** | **0.781** |
| | | *Val* | 0.695 | 0.842 | 0.759 | 0.860 | 0.914 | 0.765 |
| | Moosavi-Dezfooli *et al.* [19] | *Train* | 0.672 | 0.854 | **0.771** | **0.894** | 0.900 | 0.779 |
| | | *Val* | 0.670 | 0.849 | 0.767 | 0.886 | 0.901 | 0.771 |
| $\ell_\infty$ | UAN | *Train* | 0.649 | 0.832 | **0.753** | **0.849** | **0.889** | **0.773** |
| | | *Val* | **0.666** | **0.851** | 0.750 | 0.846 | 0.881 | 0.771 |
| | Moosavi-Dezfooli *et al.* [19] | *Train* | 0.599 | 0.763 | 0.684 | 0.836 | 0.888 | 0.750 |
| | | *Val* | 0.572 | 0.760 | 0.679 | 0.823 | 0.879 | 0.738 |
| | Mopuri *et al.* [21] | *Train* | 0.219 | 0.374 | 0.356 | 0.407 | 0.370 | 0.336 |
| | | *Val* | 0.201 | 0.365 | 0.341 | 0.411 | 0.369 | 0.337 |

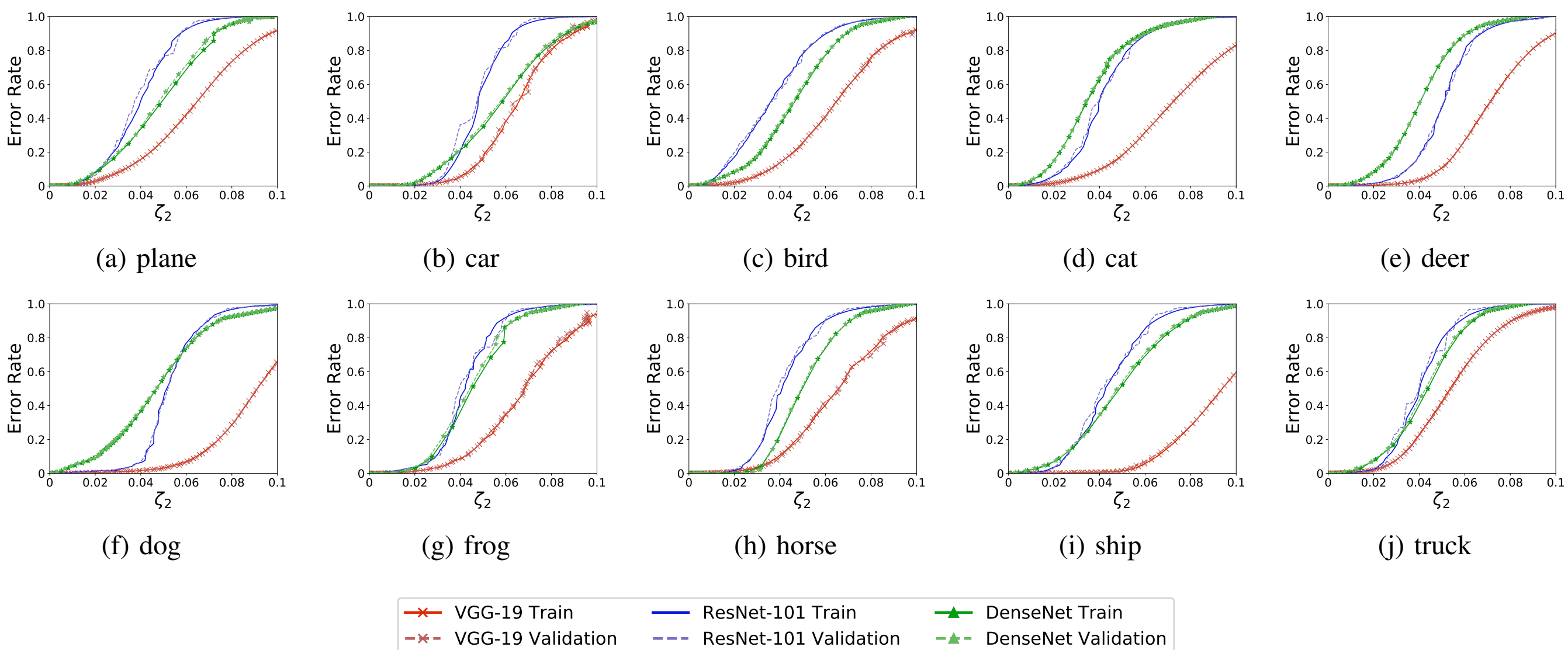

(a) plane (b) car (c) bird (d) cat (e) deer

(f) dog (g) frog (h) horse (i) ship (j) truck

Fig. 4: CIFAR-10 $\ell_\infty$ targeted attack. Each figure shows the error rate as the size of the adversarial perturbation is increased. This can be interpreted as the success rate of fooling the target model into classifying any image in CIFAR-10 as the chosen class.

TABLE IV: Error rates for non-targeted CIFAR-10 attack, under the $\ell_\infty$ metric. UAPs are constructed using row models and tested against pre-trained column models.

| | VGG-19 | DenseNet | ResNet-101 |
|---|---|---|---|
| VGG-19 | 0.666 | 0.550 | 0.612 |
| DenseNet | 0.543 | 0.750 | 0.648 |
| ResNet-101 | 0.514 | 0.681 | 0.851 |
| Ensemble | 0.499 | 0.742 | 0.849 |

At the beginning of training, there is litle structural similarity between $\mathcal{U}(z_1)$ and $\mathcal{U}(z_2)$. Throughout training the SSIM score never increases beyond 0.8, while the MSE continually increases. While the structural similary of UAPs learned by a UAN is high, it does learn to generalize to multiple UAPs that are unique from one another. Similar effects, albeit scaled down due to the smaller image size, are found for the CIFAR-10 dataset in Figure 10.

Does a UAN that learns to generalize to multiple UAPs do so to the detriment of attack accuracy? We verify this is not the case by training a UAN on a fixed noise vector and comparing to a UAN trained with non-fixed noise vectors. We found similar error rates for the two settings (see Table V); there is no loss in accuracy by extending a UAN to output multiple adversarial perturbations.

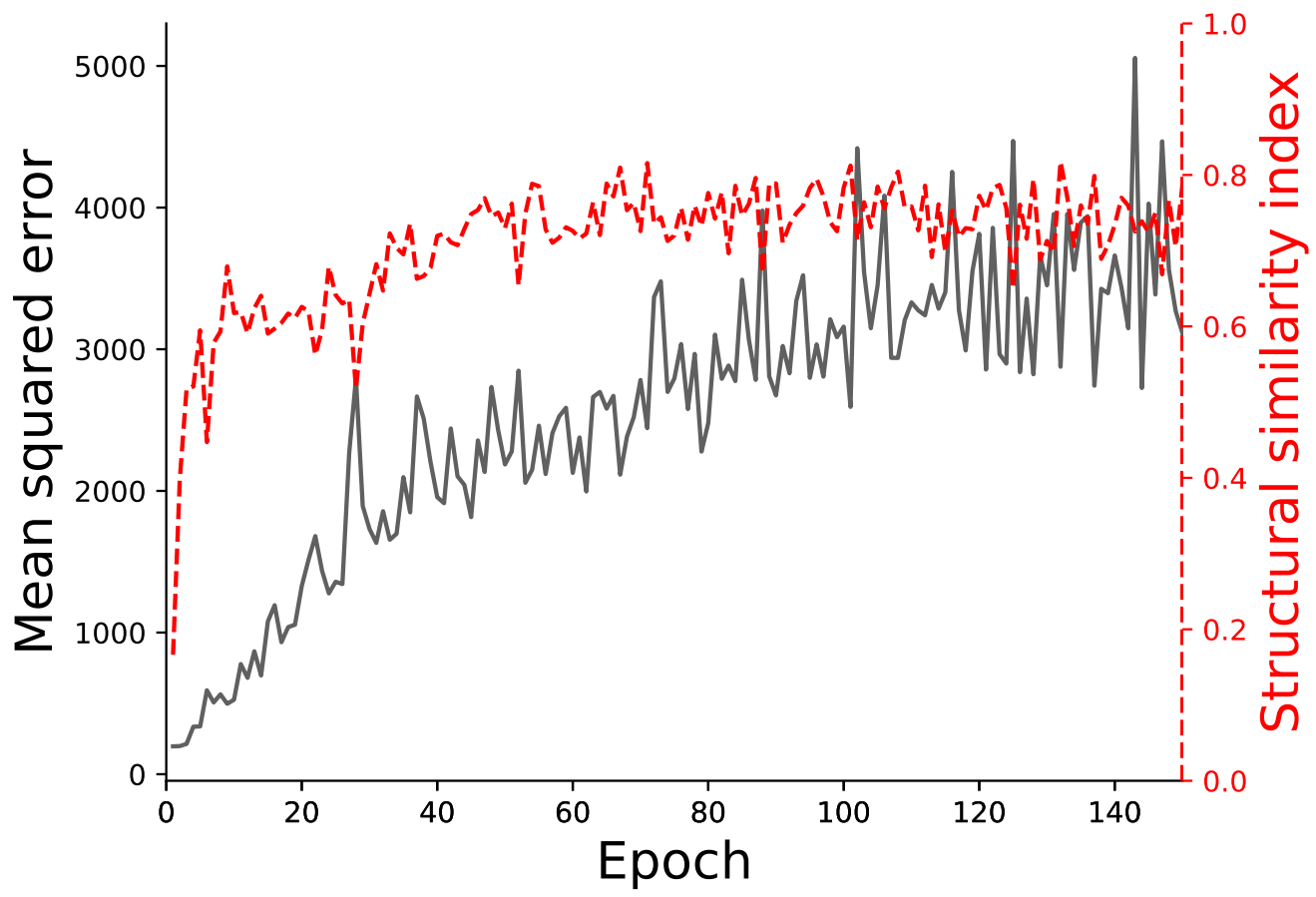

Fig. 5: MSE and SSIM scores of UAPs throughout training a UAN against VGG-19 for the ImageNet dataset.

TABLE V: Error rates for $\ell_\infty$ attacks on CIFAR-10. We compare between a UAN trained on fixed noise vectors and a UAN trained on non-fixed noise vectors.

| | Fixed $z$ | Non-fixed $z$ |
|---|---|---|
| VGG-19 | 0.661 | 0.666 |
| RESNET-101 | 0.859 | 0.851 |
| DENSENET | 0.760 | 0.750 |

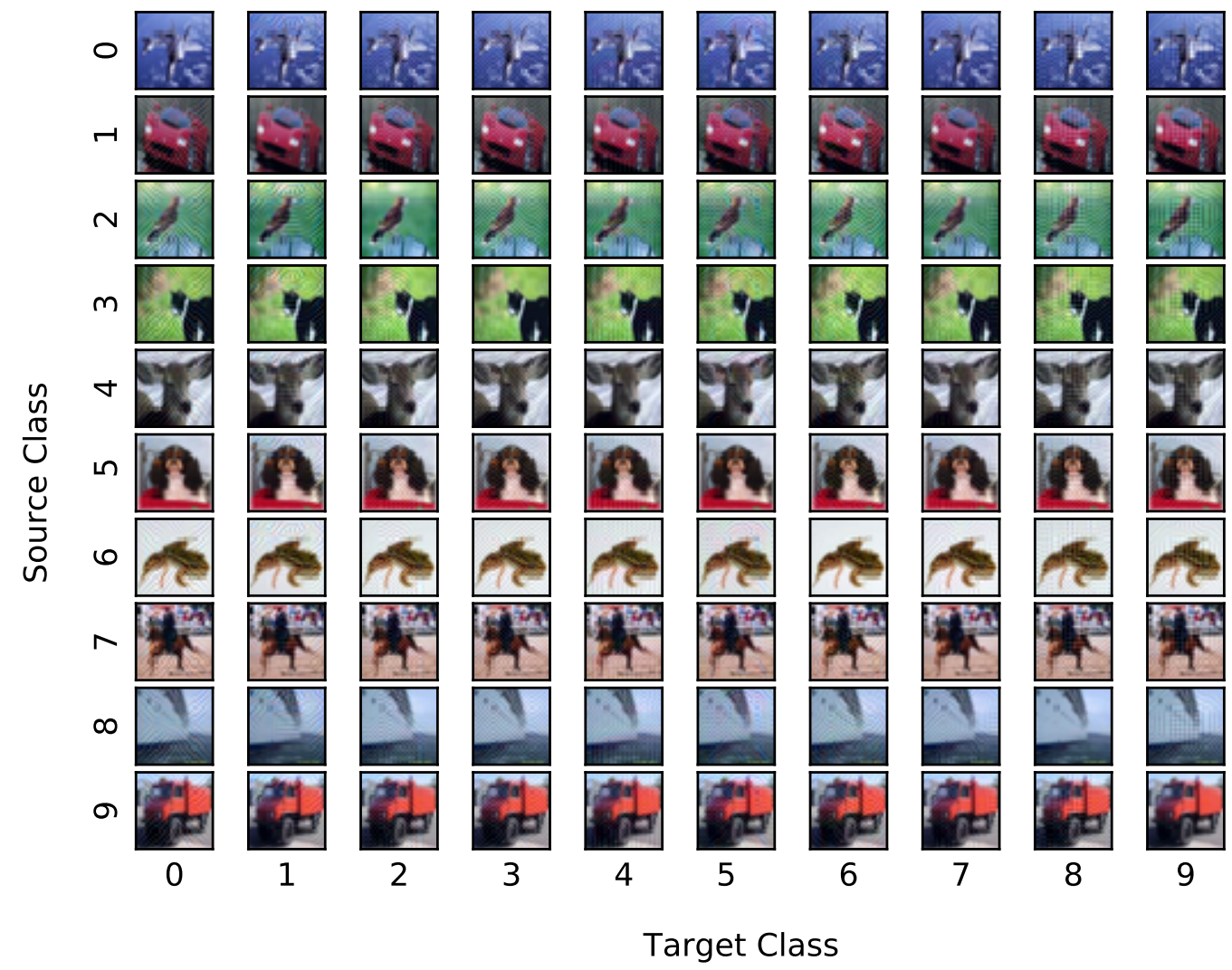

Fig. 6: Our $\ell_\infty$ attack against a DenseNet target model on the CIFAR-10 dataset, for every source/target pair. Displayed images were selected at random.

### D. Targeted Attacks

We follow the same experimental set-up as in Section IV-A, however now the attacker chooses a class, $c$, they would like the target model to classify an adversarial example as, and success is calculated as the probability that an adversarial example is classified as $c$. Figure 4 shows, for each class in CIFAR-10, the error rate of the target model as we allow larger perturbations. For nearly every class, attacks on ResNet-101 are most successful, while attacks on VGG-19 are least successful. This is in agreement with our findings in a non-targeted attack setting (cf. Table III). Despite VGG-19 being the most difficult target model to attack, it is the most well calibrated; the error rate on the training set is nearly identical to the error rate on the validation set for all classes, while there are small deviations between these two scores for ResNet-101 and DenseNet.

By looking only at results on VGG-19, one may infer that the choice of target class heavily influences the error rate (e.g. crafting UAP's for the dog and ship classes is more difficult than others). However, this is not replicated with ResNet-101 or DenseNet. We do not observe any dependencies between attack success and the target class; the attack success at different perturbation rates is similar for all classes. Figure 6 shows this attack applied to a DenseNet target model for the CIFAR-10 dataset for all source/target class pairs. Nearly all attacks are indistinguishable from the source image. Similar results are found in Figure 11 and Figure 12 for VGG-19 and ResNet-101 target models, respectively.

Interestingly, all targeted attacks follow a sigmoidal curve shape. Empirically, we found that for all three target models, there existed images that were *weakly classified* correctly (there was almost no difference between the largest probability score and probability score at the target class) and *strongly classified* correctly (there was three to four orders of magnitude difference between the probability score at the largest class and the probability score at the target class). At the beginning of training, the UAN discovers a perturbation that causes misclassifications when applied to the weakly classified images, but takes longer to find adversarial perturbations for the majority of images, resulting in a long tail at the beginning of training. With a similar effect taking place at the end of training to find adversarial perturbations for strongly classified images.

For the ImageNet dataset, we selected three classes at random and performed a targeted attack. Error rates and selected samples are given in Figures 13 to 15. We observed that the generated UAPs resembled the structure of the target class. For example, a golf ball pattern can be clearly seen in perturbations in Figure 13.

### E. Importance of training set size

So far, we have assumed the attacker shares full access to any images that were used to train the target model. However in practice, this may not be the case - an attacker may only have access to the type or a subsample of the training data. We therefore evaluate our non-targeted $\ell_\infty$ attack under stronger assumptions of attacker access to training data.

Figure 7 shows the error rate caused by a UAN trained on subsets of the CIFAR-10 training set. As expected, training on more data samples improves the success of the attack; perturbations from a UAN trained on only 50 images (5 from each class) fools 17.1% of validation set images in ResNet-101. The attack is successful when applied to nearly a fifth of images while only learning from 0.1% of the training set. The attack succeeds in 80.2% of cases when trained on 20% of the training set - in other words, there is virtually no difference in test accuracy when training on between 80-100% of the training set.

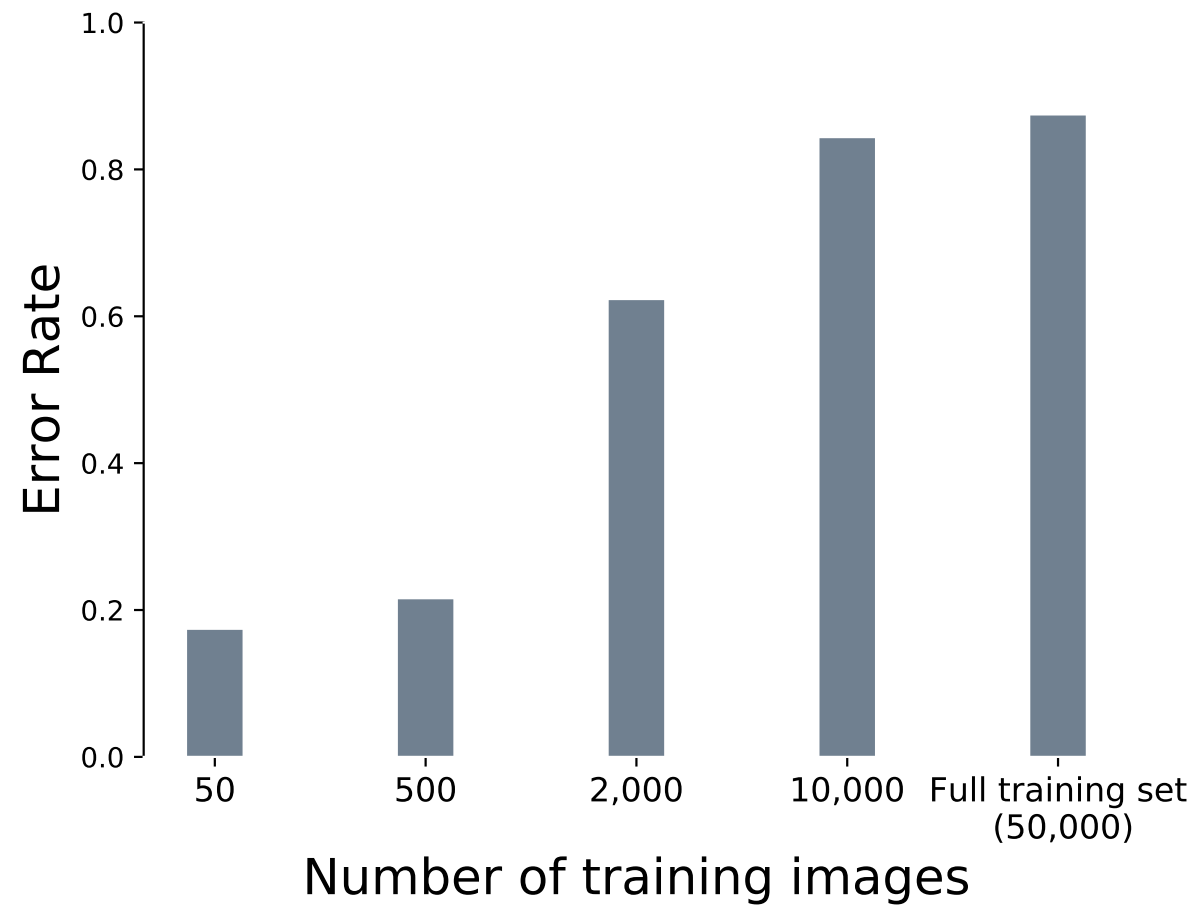

Fig. 7: Non-targeted $\ell_\infty$ attack against ResNet-101 on the CIFAR-10 dataset. We vary the number of samples the UAN is trained on, and report results on the validation set.

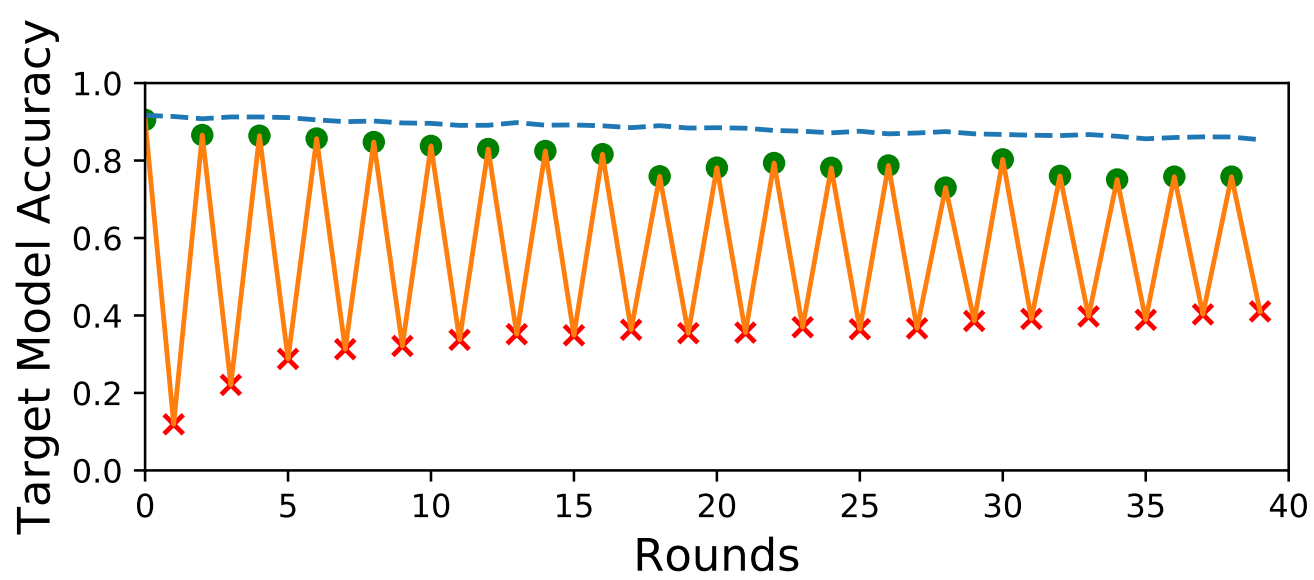

Fig. 8: A cat-and-mouse game of non-targeted $\ell_\infty$ attacks and adversarial training for a VGG-19 target model on CIFAR-10. The upper green points are the target model accuracies on adversarial images after adversarial training, the lower red crosses are the target model accuracies on adversarial images after the attack. The dotted line is target model accuracy on source images.

We find no significant difference in error rates between a UAN that has been trained on many data samples and few data samples. The amount of data samples provided to the UAN does not significantly impact its ability to learn to craft adversarial perturbations, all that must be known is the structure of the dataset on which the target model was trained. We note that this is in agreement with Papernot *et al.'s* [24] findings on the number of source images required to launch attacks on black-box models.

In addition to measuring attacker success for different training set sizes, we experimented with different batch sizes, ranging from 16 to 128, for the CIFAR-10 dataset. However, we did not observe any significant deviations in the error rate.

## V. Attacking Adversarial Training

Adversarial training [7], [16] modifies the training of a model in order to make it more robust to adversarial examples. During training, the loss function $L(\theta, x, y)$ is replaced by $\alpha \cdot L(\theta, x, y) + (1-\alpha) \cdot L(\theta, x+\delta', y)$. By augmenting the original data to include adversarial counterparts, the model learns to classify adversarial examples correctly. Non-generative attacks have shown to be successful against adversarially trained models, however, recent work [1] suggested that this may not be the case for UAPs. In [1], adversarial training is successfully applied to a CIFAR-10 classifier, effectively eliminating the adversarial effect of UAPs.

In our work, we verified that this is case; adversarial training eliminates UAP success. However, we find that adversarially trained models are still vulnerable to UAN trained against the defended model.

Similarly to Hamm [8], we play a cat-and-mouse game where (1) a UAN is trained against a target model, and (2) the target model is retrained with adversial examples crafted from (1) (denoted ADV TM). This generates a sequence: UAN1 $\rightarrow$ ADV TM1 $\rightarrow$ UAN2 $\rightarrow$ ADV TM2 $\rightarrow$ UAN3 $\rightarrow$ .... We let this game play out for many rounds, and claim that if adversarial training is a defense against UAPs, over many rounds the classification error on adversarial examples should tend to zero.

Figure 8 shows such a cat-and-mouse game over 20 rounds of (1) and 20 round of (2). An adversarially trained target model is able to classify nearly all adversarial examples correctly, at any given round. However, attacks against adversarially retrained models are only somewhat mitigated; there is a 25% reduction is attack success between the first and final round. After this, the cycle reaches an equilibrium, with no improvement in successive attacks or defended models.

## VI. Conclusion

We presented a first-of-its-kind universal adversarial example attack that uses machine learning at the heart of its construction. We comprehensively evaluated the attack under many different settings, showing that it produces quality adversarial examples capable of fooling a target model in both targeted and non-targeted attacks. The attack transfers to many different target models, and improves on other state-of-the-art universal adversarial perturbation construction methods.

## VII. Acknowledgements

Jamie Hayes is funded by a Google PhD Fellowship in Machine Learning.

TABLE VI: Error rates for non-targeted $\ell_\infty$ attacks on ImageNet.

| | VGG-19 | INCEPTION-V1 [33] |
|---|---|---|
| UAN | 0.846 | 0.809 |
| Poursaeed *et al.* [26] | 0.801 | 0.792 |
| Mopuri *et al.* [27] | 0.838 | 0.904 |

## APPENDIX A
## A NOTE ON RECENT CONCURRENT WORK

We are unaware of any previous work that studies the relationship between generative models and universal adversarial perturbations. However, we note that two recent studies [26], [27] also craft perturbations using generative models [4]. Poursaeed *et al.* [26] have a similar set-up to our attack, optimizing under both $\ell_2$ and $\ell_\infty$ metrics, however, they did not include a distance minimization term within the objective function, instead relying a scaling factor before applying the perturbation to an image. The Mopuri *et al.* [27] attack is only optimized using the $\ell_\infty$ metric. They also eschew a distance minimization term, and instead include a diversity term within the objective function, so that the objective does not get stuck in a local minima resulting in a limited number of effective perturbations. We were unable to obtain source code for Mopuri *et al.'s* [27] attack and were unsuccessful in replicating results, and so we report comparison in results only on the target models that were shared in both pre-prints, on ImageNet (since other works only report results on this dataset for the task of generating UAPs) using the $\ell_\infty$ metric (see Table VI).

[4]Poursaeed *et al's.* [26] pre-print was made available online 12 hours before our first version was made available, while Mopuri *et al's.* [27] pre-print was made available three days afterwards.

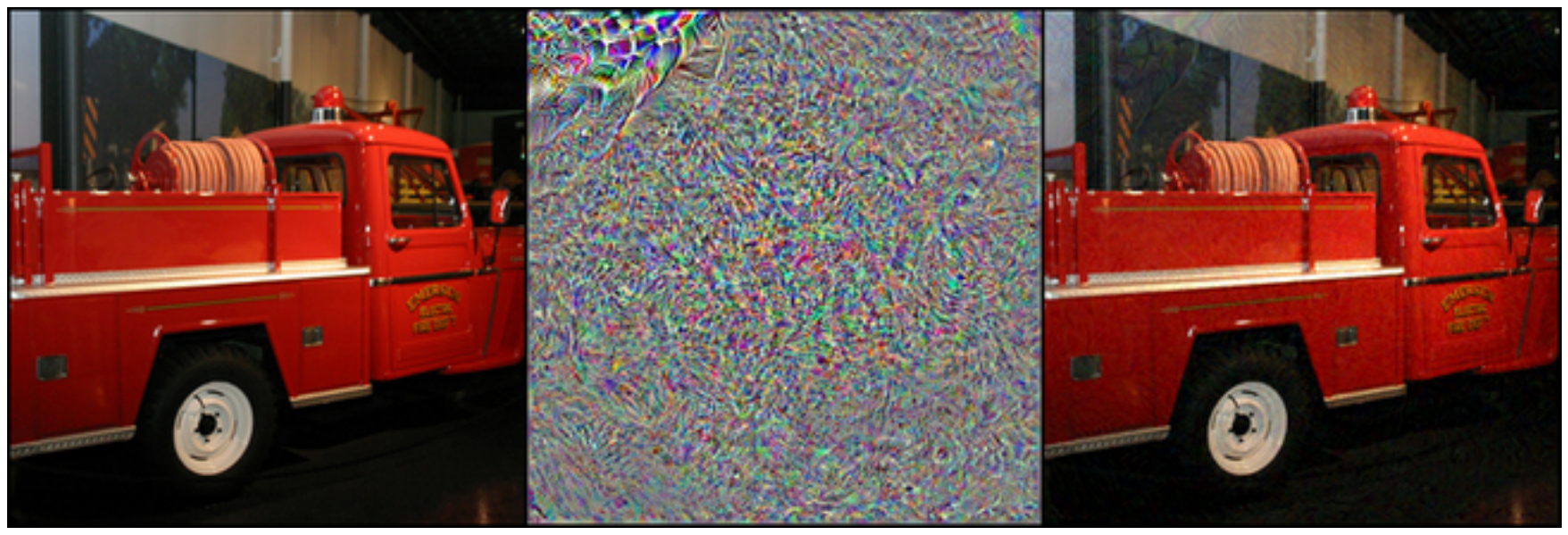

(a) Inception-V3: Fire engine (54.6%), $\delta'$, Wrecker (79.4%)

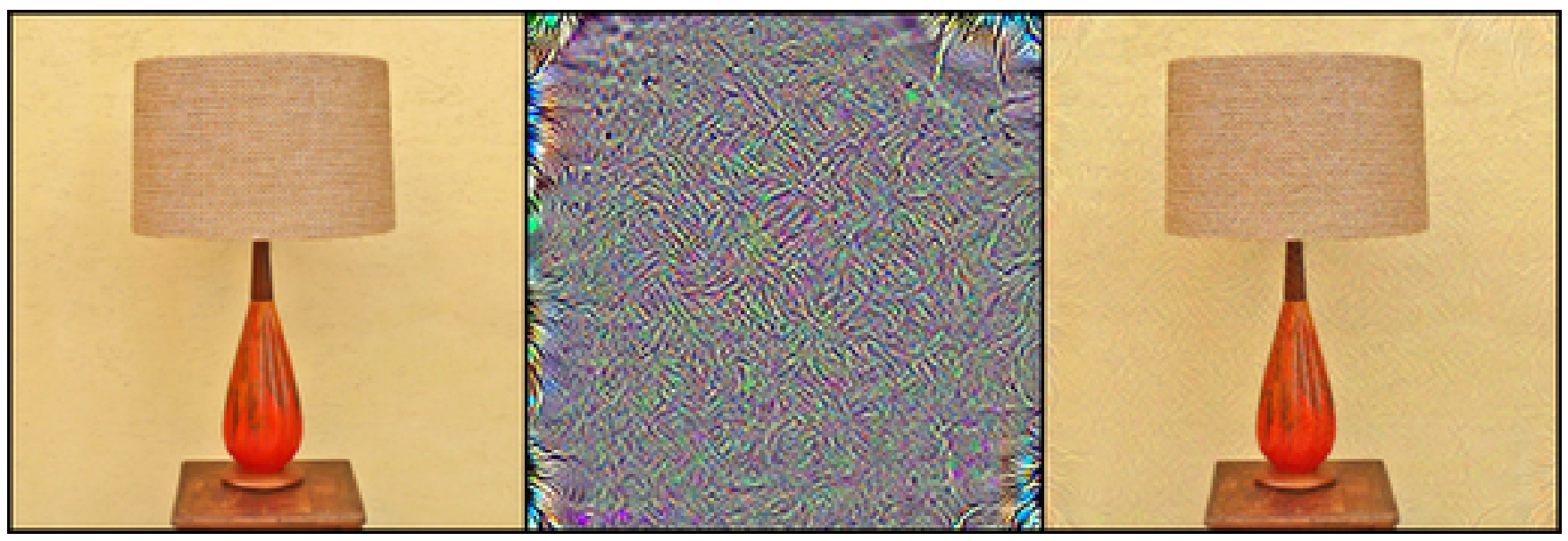

(b) ResNet-152: Table lamp (87.2%), $\delta'$, Tabby cat (41.9%)

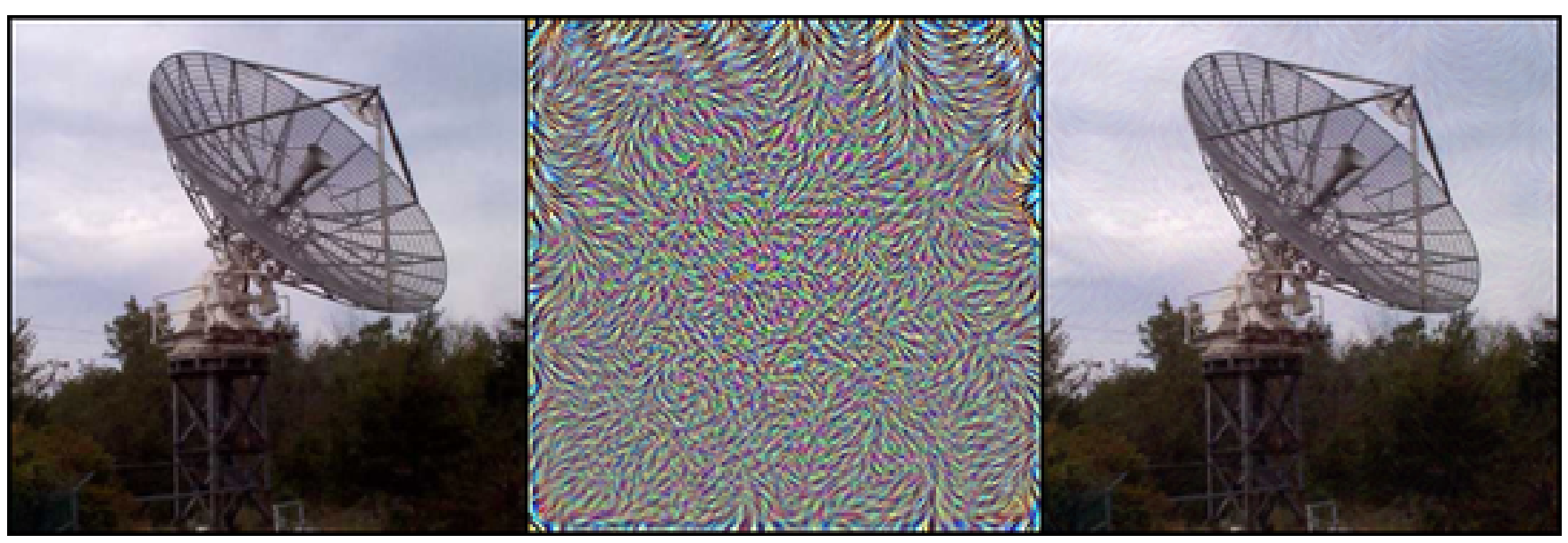

(c) VGG-19: Radio telescope (97.5%), $\delta'$, Great Pyrenees (36.7%)

Fig. 9: Selection of successful adversarial examples (with target model confidence) from non-targeted $\ell_\infty$ attacks on ImageNet. From left to right: Source image, UAP, adversarial image.

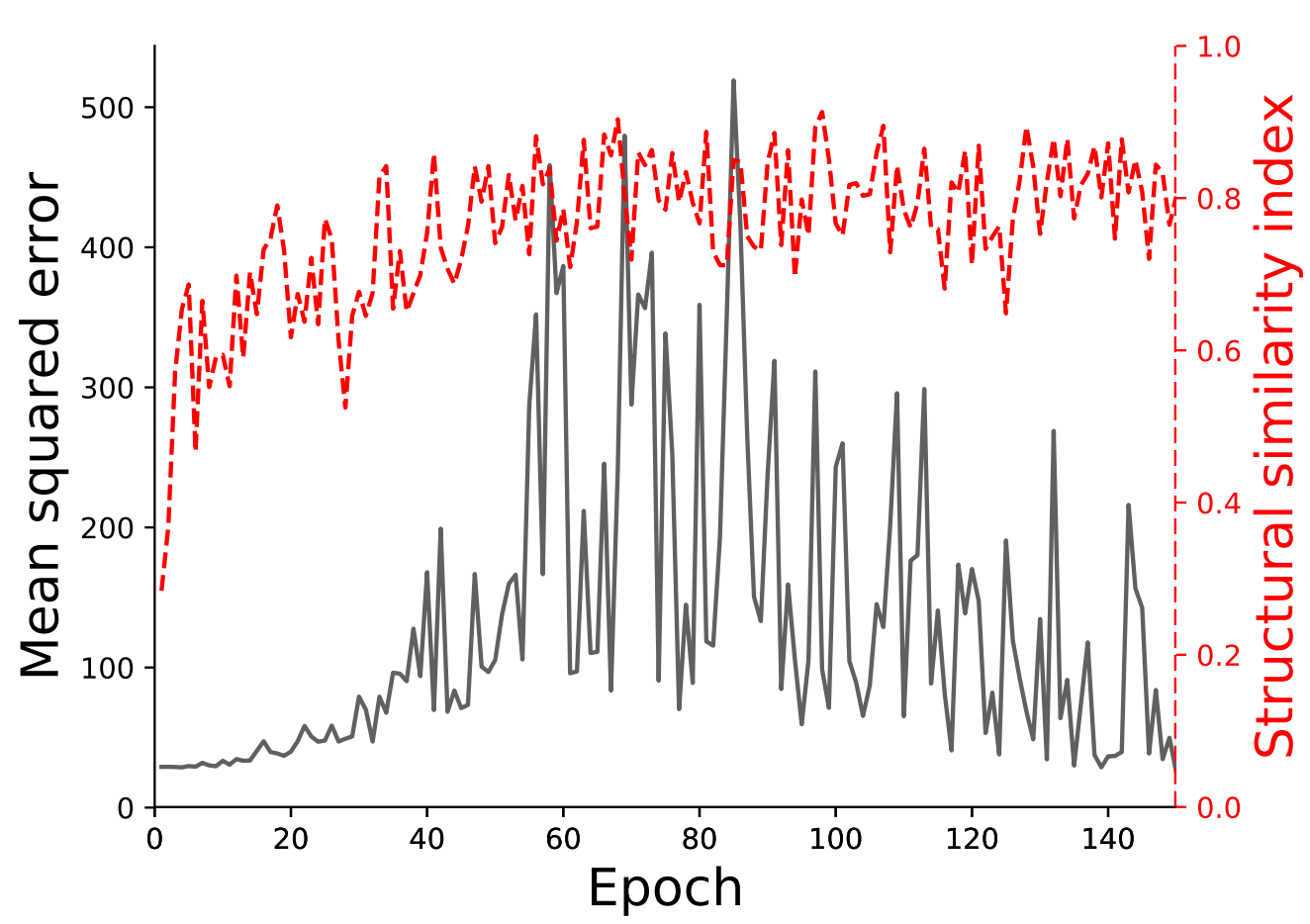

Fig. 10: MSE and SSIM scores of UAPs throughout training a UAN against VGG-19 for the CIFAR-10 dataset.

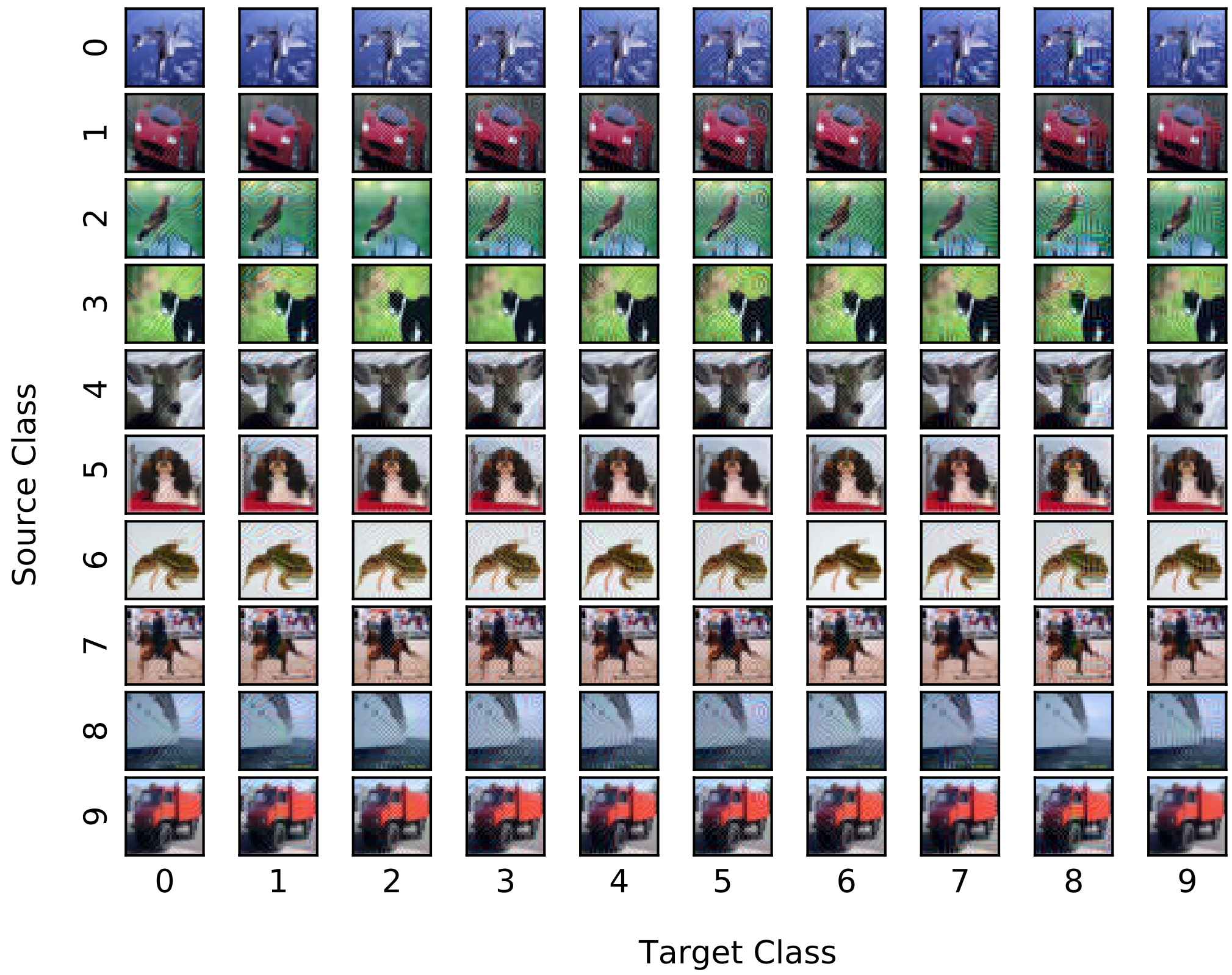

Fig. 11: Our $\ell_\infty$ attack against a VGG-19 target model on the CIFAR-10 dataset, for every source/target pair. Displayed images were selected at random.

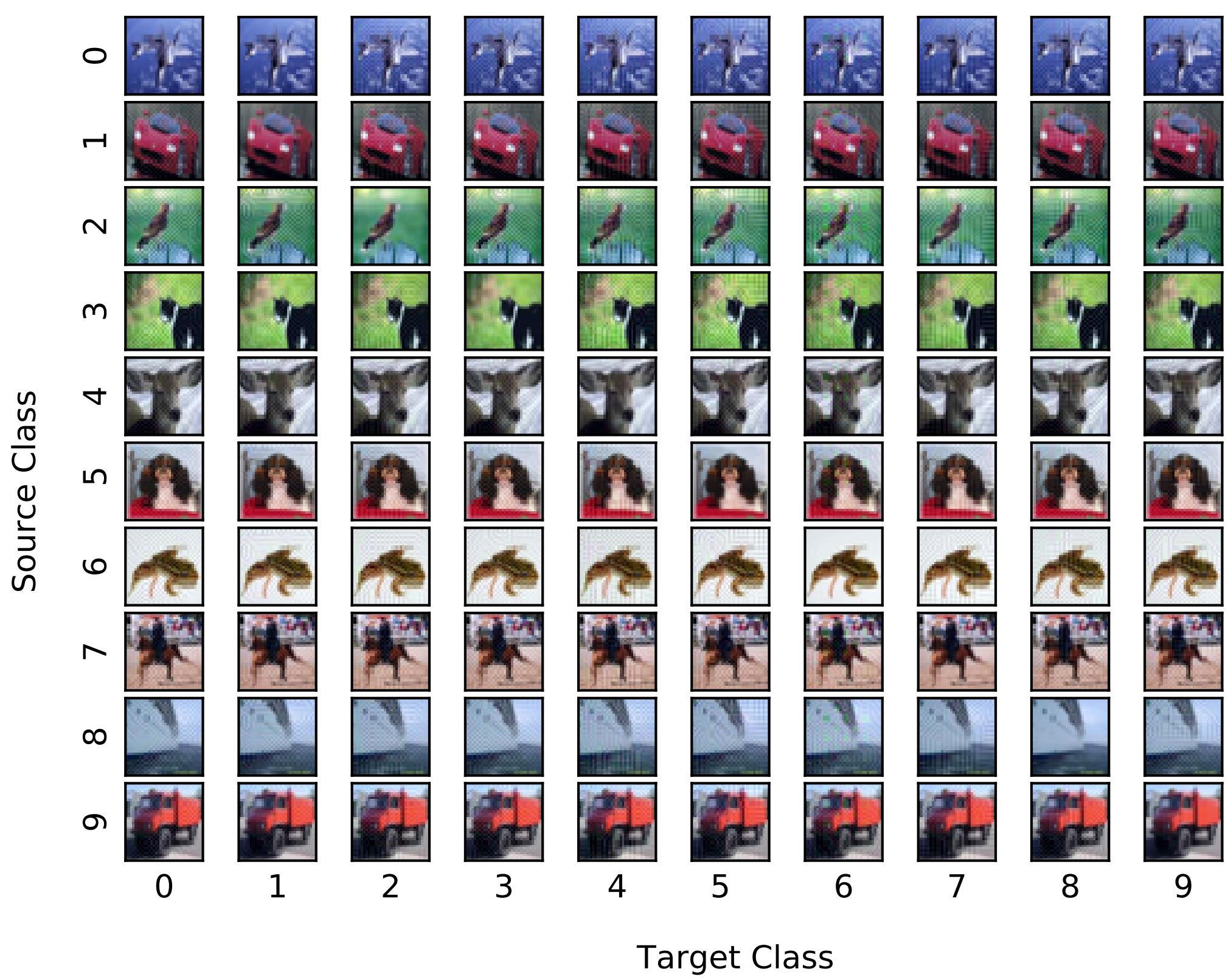

Fig. 12: Our $\ell_\infty$ attack against a ResNet-101 target model on the CIFAR-10 dataset, for every source/target pair. Displayed images were selected at random.

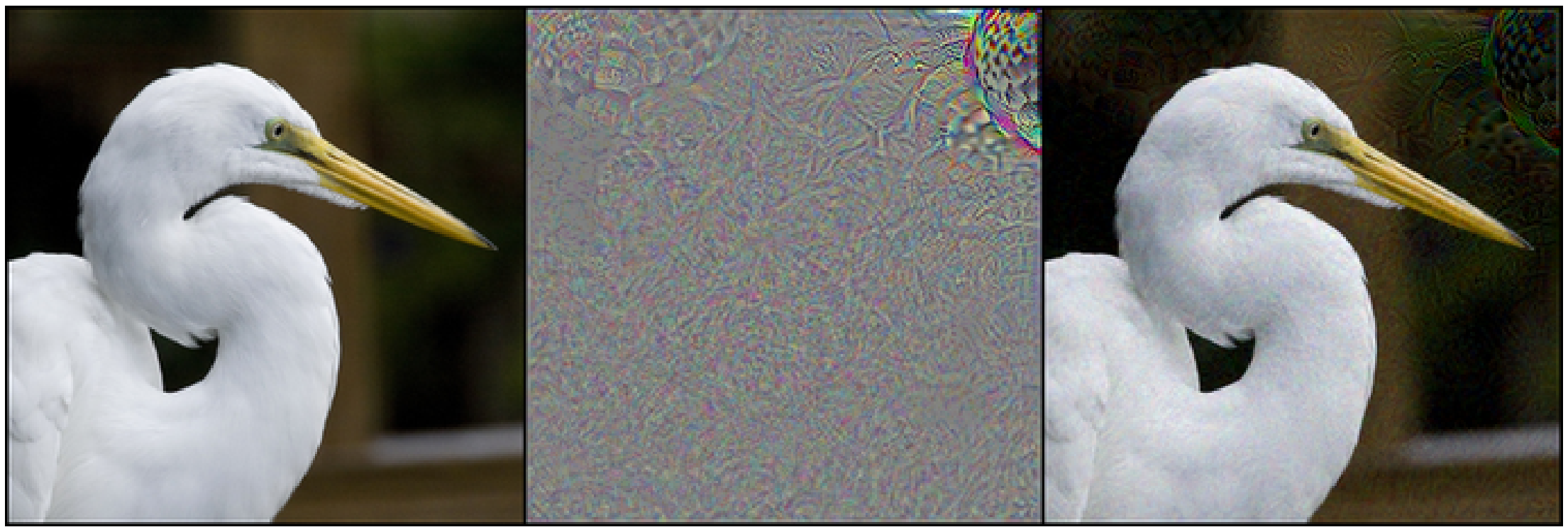

(a) Inception-V3: American egret (95.0%), $\delta'$, Golf ball (98.8%). Overall target model error rate: 0.654

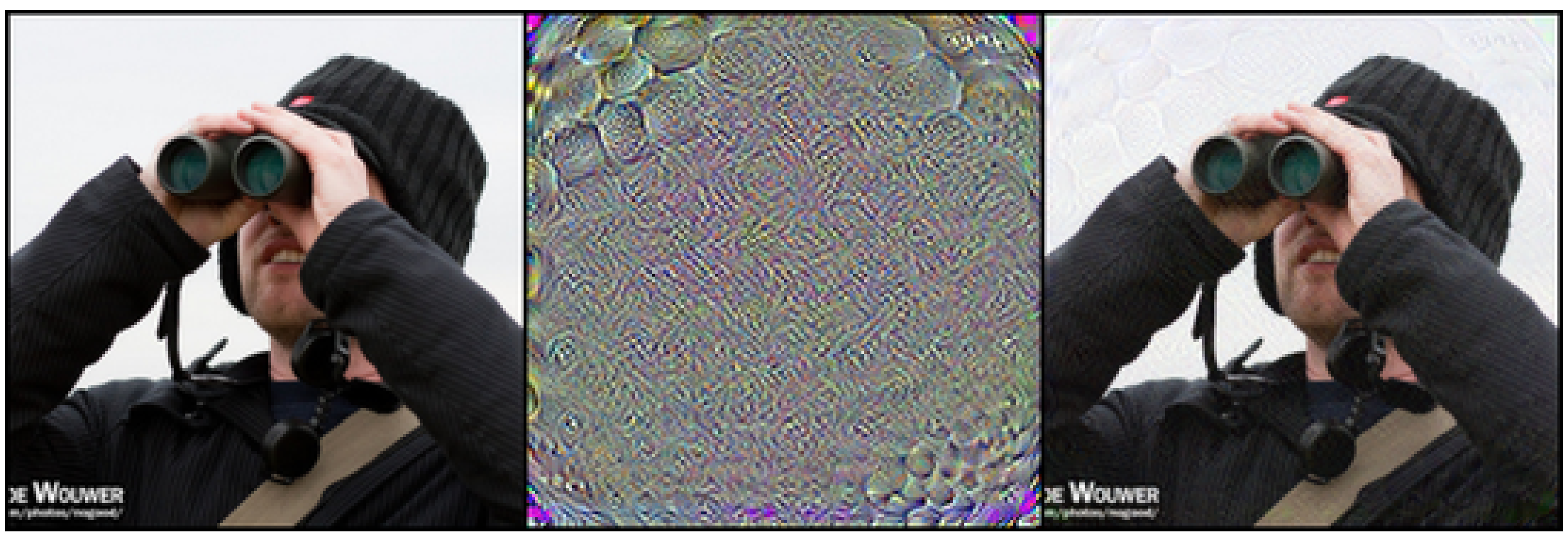

(b) ResNet-152: Binoculars (99.9%), $\delta'$, Golf ball (62.9%). Overall target model error rate: 0.734

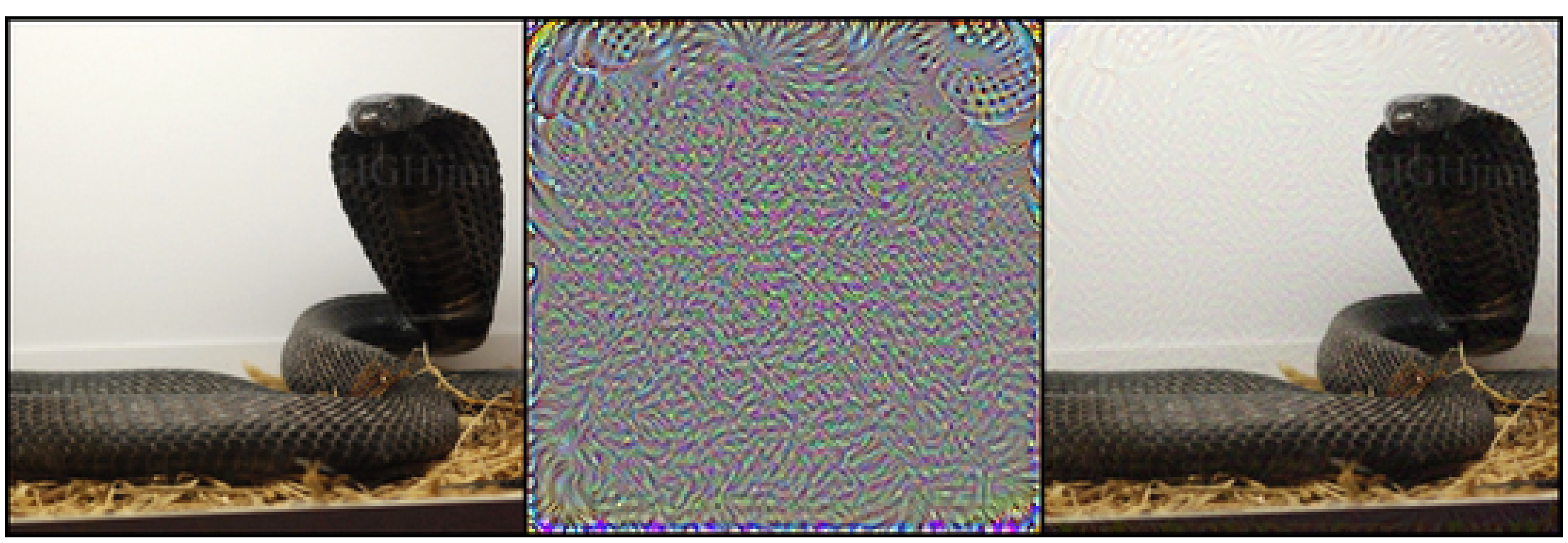

(c) VGG-19: Indian cobra (99.9%), $\delta'$, Golf ball (99.7%). Overall target model error rate: 0.514

Fig. 13: Selection of successful adversarial examples (with target model confidence) for targeted $\ell_\infty$ attacks on ImageNet. The target class was randomly chosen to be *Golf ball*. From left to right: Source image, UAP, adversarial image.

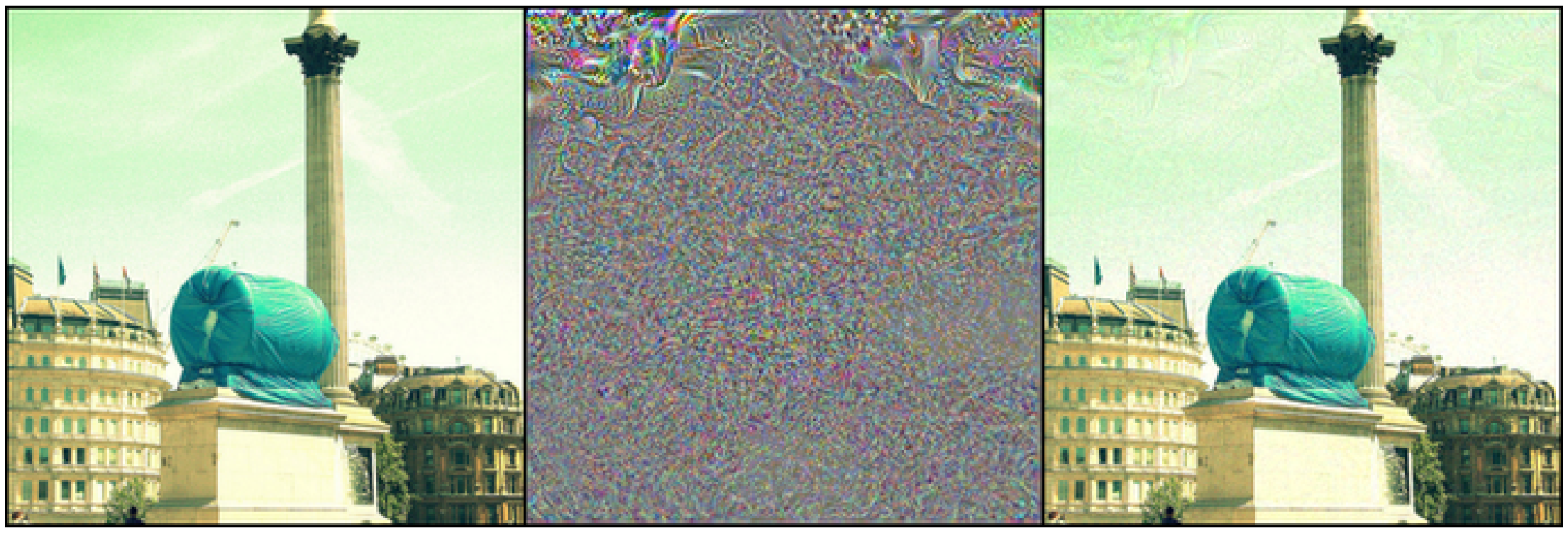

(a) Inception-V3: Pedestal (98.4%), $\delta'$, Broccoli (88.7%). Overall target model error rate: 0.598

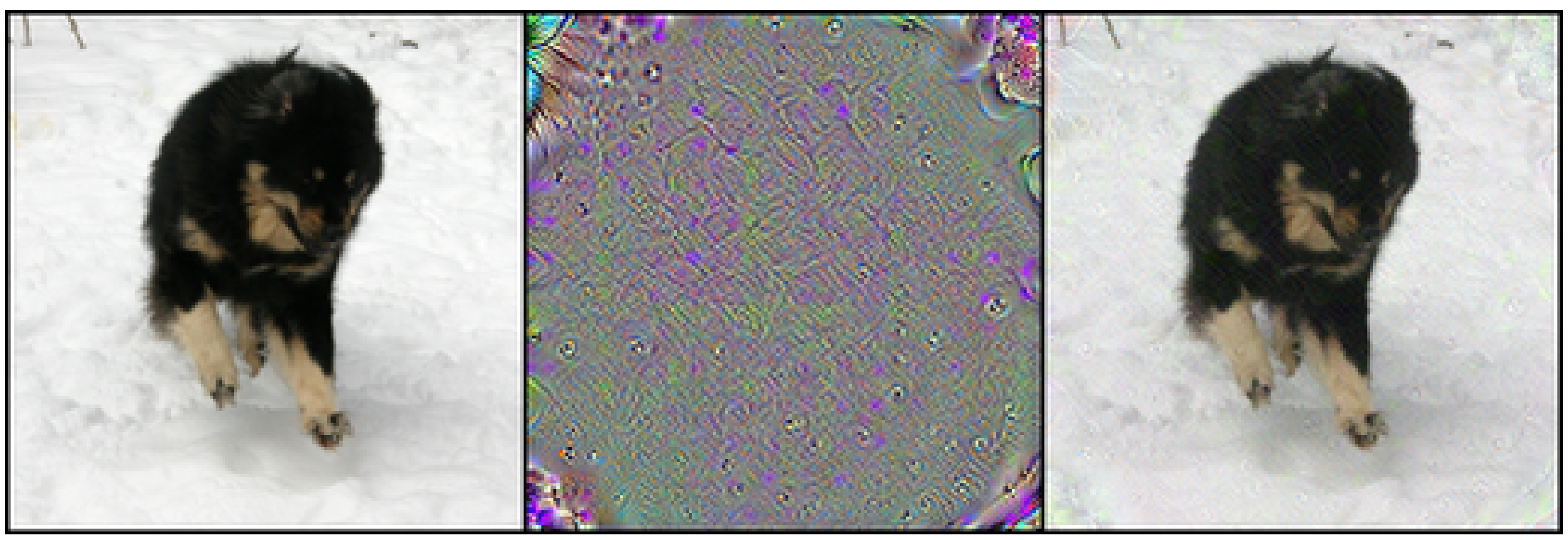

(b) ResNet-152: Tibetan mastiff (88.4%), $\delta'$, Broccoli (98.1%). Overall target model error rate: 0.691

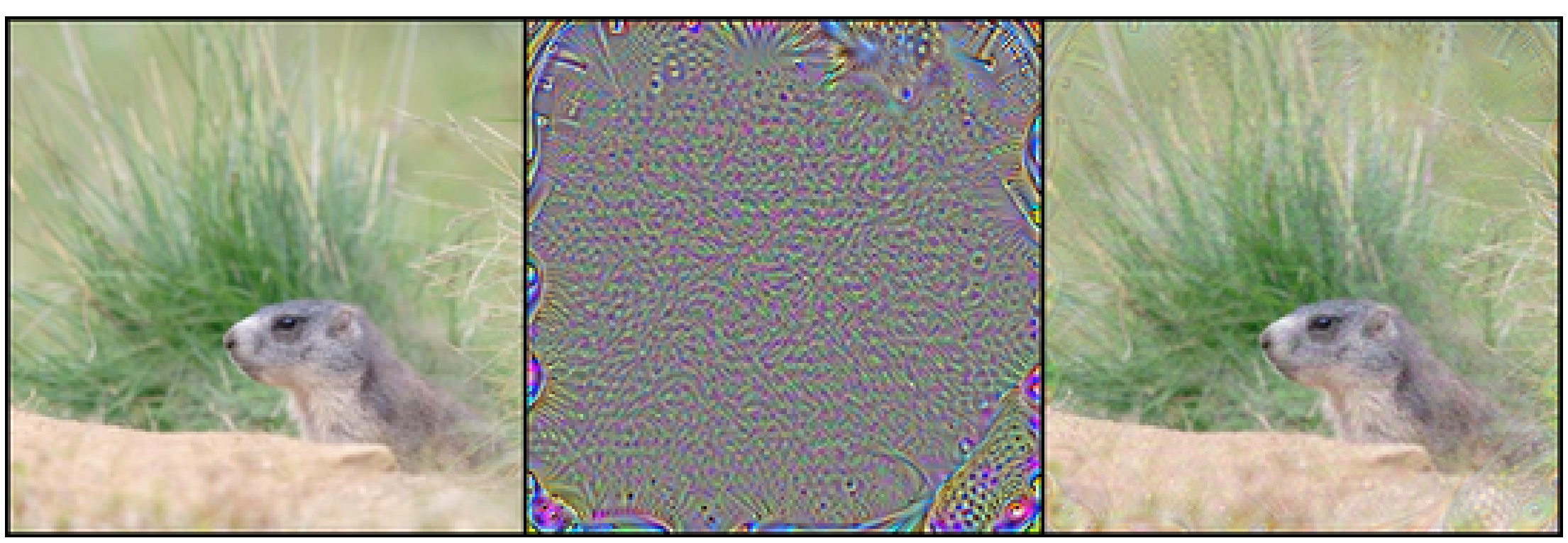

(c) VGG-19: Marmot (95.4%), $\delta'$, Broccoli (48.4%). Overall target model error rate: 0.480

Fig. 14: Selection of successful adversarial examples (with target model confidence) for targeted $\ell_\infty$ attacks on ImageNet. The target class was randomly chosen to be *Broccoli*. From left to right: Source image, UAP, adversarial image.

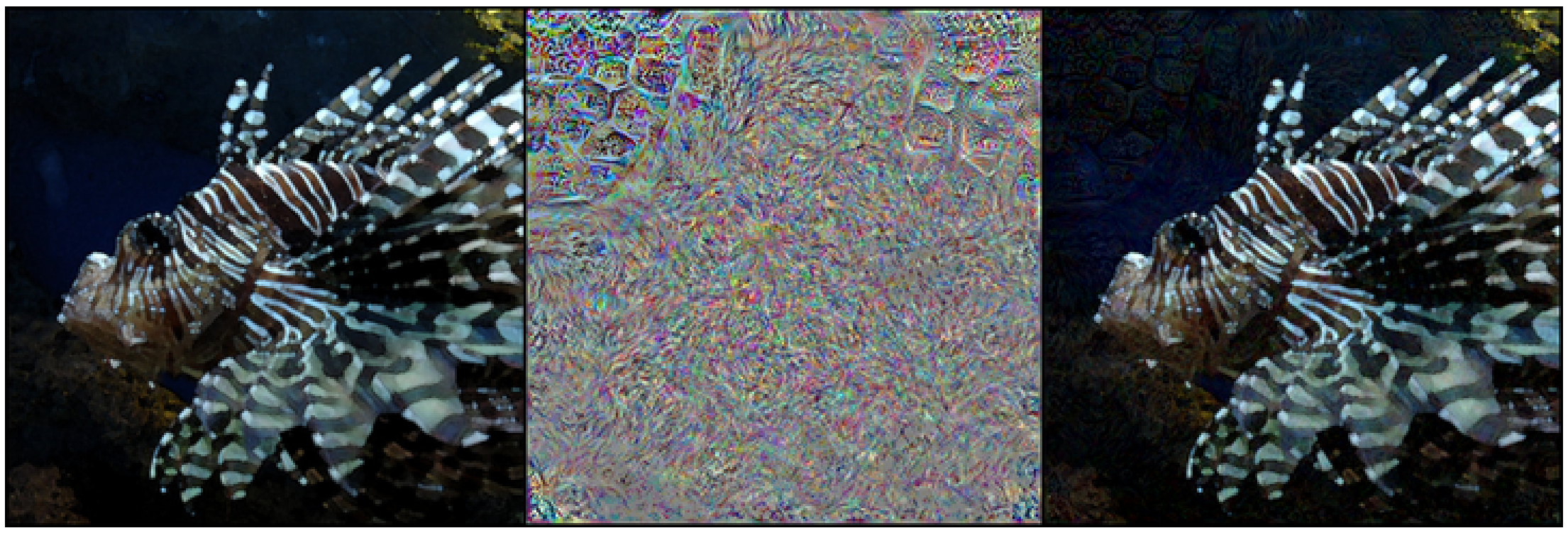

(a) Inception-V3: Lionfish (89.7%), $\delta'$, Stone wall (54.0%). Overall target model error rate: 0.533

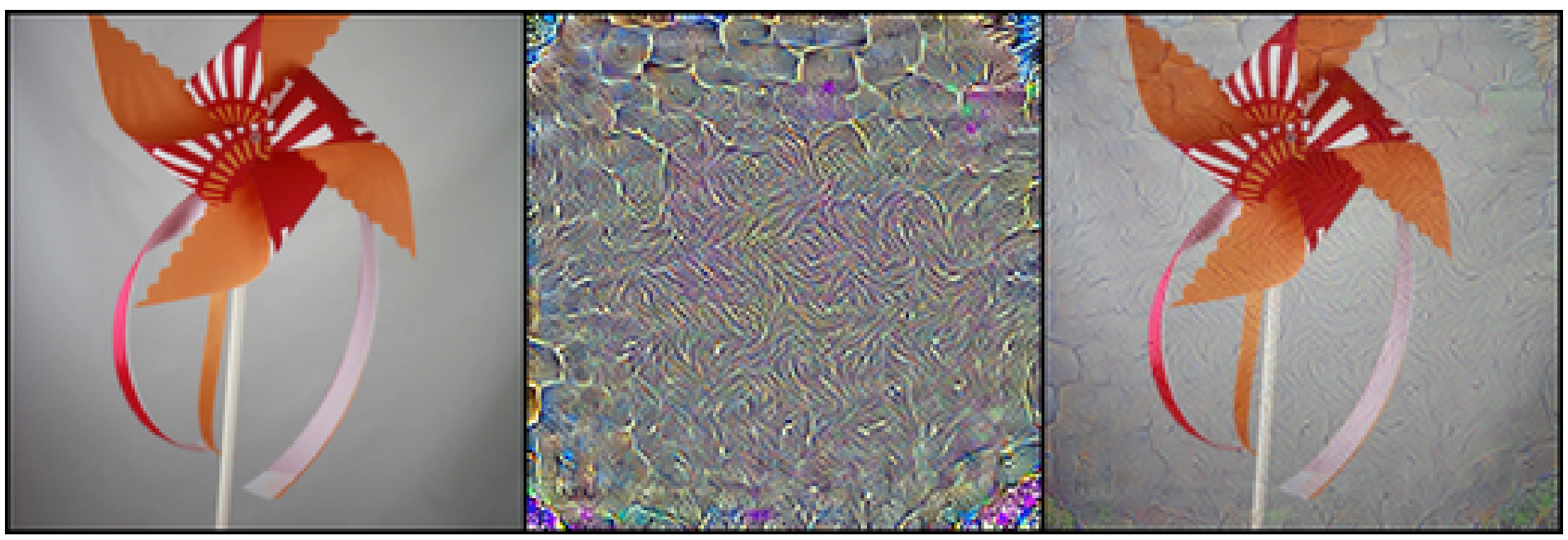

(b) ResNet-152: Pinwheel (99.9%), $\delta'$, Stone wall (47.0%). Overall target model error rate: 0.587

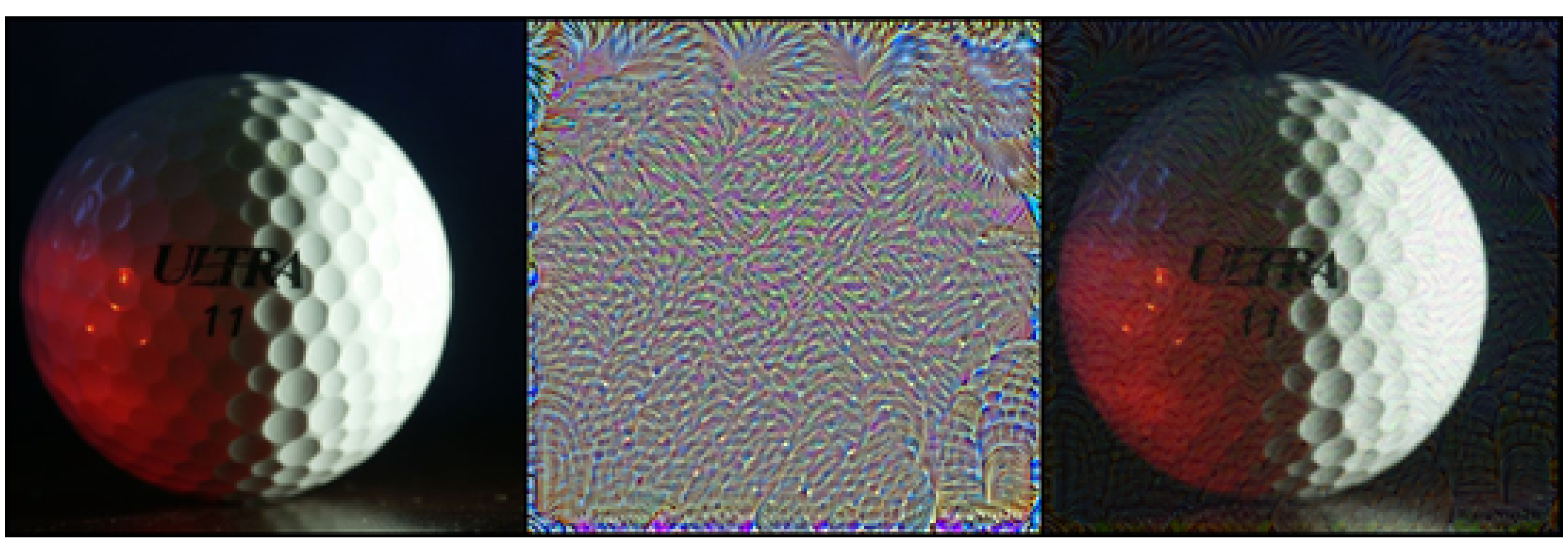

(c) VGG-19: Golf ball (99.9%), $\delta'$, Stone wall (23.7%). Overall target model error rate: 0.447

Fig. 15: Selection of successful adversarial examples (with target model confidence) for targeted $\ell_\infty$ attacks on ImageNet. The target class was randomly chosen to be *Stone wall*. From left to right: Source image, UAP, adversarial image.